# LSST: a Complementary Probe of Dark Energy

J. A. Tyson and D. M. Wittman[a], J. F. Hennawi and D. N. Spergel[b]

[a]Bell Labs, Lucent Technologies,
600 Mountain Ave., Murray Hill, NJ 07974

[b]Astrophysical Sciences, Princeton University,
Princeton, NJ 08544

The number of mass clusters and their distribution in redshift are very sensitive to the density of matter $\Omega_m$ and the equation of state of dark energy $w$. Using weak lens gravitational tomography one can detect clusters of dark matter, weigh them, image their projected mass distribution, and determine their 3-D location. The degeneracy curve in the $\Omega_m - w$ plane is nearly orthogonal to that from CMB or SN measurements. Thus, a combination of CMB data with weak lens tomography of clusters can yield precision measurements of $\Omega_m$ and $w$, independently of the SN observations. The Large Synoptic Survey Telescope ($LSST$) will repeatedly survey 30,000 square degrees of the sky in multiple wavelengths. $LSST$ will create a 3-D tomographic assay of mass overdensities back to half the age of the universe by measuring the shear and color-redshift of billions of high redshift galaxies. By simultaneously measuring several functions of cosmic shear and mass cluster abundance, $LSST$ will provide a number of independent constraints on the dark energy density and the equation of state.

## 1. INTRODUCTION

To date, most of what we know about the large-scale structure of the universe comes from the observed anisotropies in the cosmic microwave background (CMB) and from the distribution of galaxies and supernovae. The CMB provides the earliest sample of mass fluctuations, from a time when the universe was 50,000 times younger. Different cosmological models predict different volume-redshift relations and different scenarios in the growth of mass structures over cosmic time, so comparison of the CMB-derived mass spectrum with that seen at later times will be a powerful test of cosmology. The large-scale mass distribution at late times has traditionally been characterized through the large-scale galaxy distribution, on the assumption that galaxies trace mass in a simple way. But weak gravitational lensing can trace structure more directly, is sensitive to the comoving volume, and relies on no "standard metrics." For the supernova test of the luminosity distance, there remain questions of whether Type Ia supernovae are accurate "standard candles" over the relevant range of look-back times.

$LSST$ will utilize the physics-based technique of gravitational lensing in which angles and redshifts are measured, yielding direct maps of dark mass in 3-D, unbiased to baryons and radiation. $LSST$ will measure the volume-redshift relation and the mass structure development over the range of cosmic time during which it is currently thought that the universe transitioned from matter-dominated to dark energy-dominated.

The evolution of mass clustering is the most sensitive test of our current dark energy and dark matter cosmology[10,14]. Calibrated maps of mass as a function of cosmic look-back time can (1) constrain the nature of the dark matter by its power spectrum, by the way it clumps gravitationally over time, and by its detailed distribution (voids, walls, filaments); (2) test the cosmology through cosmic shear, the cumulative shear due to all mass overdensities out to high redshift; (3) probe dark energy content in a way complementary to and entirely independently from CMB+SN, through the time evolution of the power spectrum; and (4) sharply constrain the dark energy equation of state (to about one percent) through the time evolution of the number of mass overdensities. Such maps must be obtained from statistical inversion of the observed shear of a billion high redshift background galaxies, but with existing facilities it would take hundreds of years to accumulate sufficient data to definitively address these questions. $LSST$ will address these questions within a decade.

## 2. WIDE DEEP FAST

The various direct observational tests of cosmology (weak lensing, CMB anisotropy, and SNe) each suffer from degeneracies. The CMB measurements are currently by far the most accurate of the three, but still are sensitive only to a combination of dark matter and dark energy rather than one or the other. The luminosity distance vs redshift (SNe Ia test) by itself has moderate sensitivity to the equation of state of dark energy, but with an $\Omega_m$ prior, SNe Ia observations may be used to constrain the ratio of density to pressure in the dark energy. Weak lensing measurements measure dark matter in a direct way. Each of these experiments has its strengths, and in combination they break the degeneracies. But a set of experiments which minimally breaks the degeneracies will never test the foundations.

The $\Omega_m \Omega_\Lambda$ degeneracy can also be broken by MAP CMB data, and even more cleanly later with the Planck data. The combination of weak lensing (*with* photometric redshifts) and CMB data provides a sensitive consistency test for the theory [11]. In addition the combination will constrain most of the many parameters of current theories substantially better than either alone. $LSST$ will discover many supernovae in both the time-domain search and the 1000 deg$^2$ survey (because it requires many visits to each piece of sky). Combining the Type Ia SN results with either the weak lensing or CMB results can also lift degeneracies in $\Omega_\Lambda \Omega_m$ space. Combining all three observations (weak lensing with photometric redshifts, SN, CMB) will lead to even higher precision cosmology. Comparison of different approaches will reveal systematics and lead to refinements. Most importantly, the combination will test the entire foundation of the theory. $LSST$ will produce both weak-lens and supernova probes of dark energy with the same instrument.

In addition, the large field of view opens up the time domain in search of rare, high-energy phenomena. Currently, time-domain astrophysics is limited to projects that provide shallow coverage of $\sim 10$ deg$^2$ (*e.g.* the MACHO project), or extremely shallow coverage of a larger area. With the ability to survey an entire hemisphere to 24th magnitude in a few dark nights, $LSST$ will open up new areas of parameter space. For example, faint optical bursts. Undoubtedly, other rare phenomena remain to be discovered as well.

The design of the $LSST$ is driven by the following figure of merit. In a given integration time, the size of field larger than $\theta^2$ that can be explored to given stellar magnitude is directly proportional to $A\theta^2\eta/d\theta^2$, where A is the collecting area, $\theta^2$ the solid angle of the field of view, $\eta$ the overall efficiency and $d\theta^2$ the solid angle of the seeing-limited image. Today's 8m class telescopes and detectors are superb at optimizing all of these factors except $\theta^2$ (typically $\theta^2$ is $\sim 0.04$ deg$^2$).

Advances in three areas of technology (large aspherical optics fabrication and metrology, semiconductors and terascale computation, and ULSI CCD mosaics) have come together in the design of the $LSST$ system [12]. With its large throughput and dedicated observational mode, the $LSST$ opens an unexplored region of parameter space and enables programs that would take many decades on current facilities. Previously named the "Dark Matter Telescope," the $LSST$ will be able to reach $5\sigma$ limiting surface brightness of 28-30 magnitude arcsec$^{-2}$ in the wavelength range 0.3–1 $\mu$m over a 7 deg$^2$ field in 3 nights of dark time. This opens new observational possibilities in low surface brightness wide-field surveys. For point sources, the $LSST$ will reach 24th magnitude at $5\sigma$ in only 10 seconds. Repeated imaging of large areas of the sky in this mode will probe unprecedented volumes out to high redshift in a way which offers control of image aberration systematics crucial to weak lensing.

The time required to complete the surveys described above is inversely proportional to optical throughput: square meters of collecting area times solid angle on the sky. The $LSST$ with an 8.4m primary and 7 deg$^2$ per exposure will have a throughput 50 times larger than current 4m telescopes with their large CCD mosaics. The three-mirror design will produce unprecedented image quality over the full 7 deg$^2$ field. A single 10s exposure will go to 24th magnitude ($5\sigma$). The dark mass/energy survey would utilize deep ($\sim 27$ mag) multi-color imaging over 30,000 deg$^2$ with photometric redshifts for source galaxies out to z=3.

A team has been working on $LSST$ science and design for two years (see http://lsst.org). While there are some technology challenges, no show-stoppers have been found. The optics, mechanics, control, CCD array and camera could be built today.

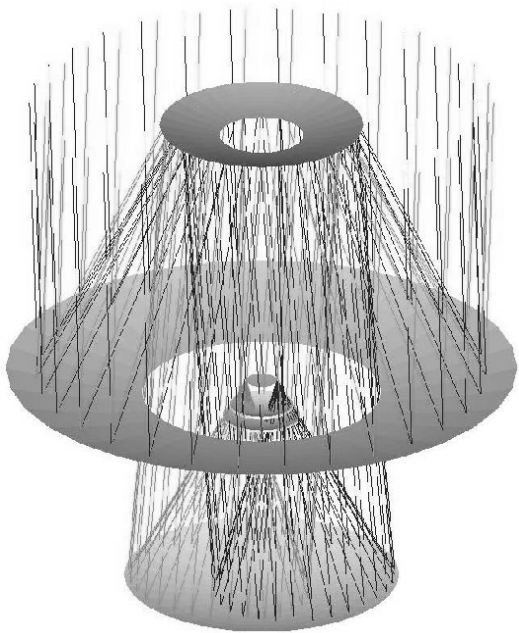

Figure 1. The LSST ray diagram. Three aspherical surfaces deliver excellent image quality over a wide field. The 2 Gpixel camera will produce a deep image of 7 square degrees in 10 seconds.

## 3. COSMIC SHEAR

One way to explore the nature of dark matter, which $LSST$ will exploit like no other experiment, is to study its distribution on large scales. This distribution tells us not only about the nature of the dark matter, but also about the global cosmological parameters that describe the universe. In addition, the evolution of this distribution with cosmic time, which $LSST$ will measure exquisitely, is a critical test of the entire framework.

In the presence of foreground mass overdensities, the light rays from galaxies narrowly separated on the sky travel similar paths past intervening mass concentrations and thus undergo similar image distortions. The resulting correlation of distant galaxy ellipticities is highest at small angular separation and drops for widely separated galaxies whose light bundles travel through completely different structures.

Because the typical background galaxy has an intrinsic ellipticity of roughly 30%, many thousands of source galaxies must be averaged together to detect this small signal. In addition, a large area of sky must be covered, because mass structures should span a few arcminutes to a degree at a typical mid-path distance of redshift $\sim 0.5$.

Driven by advances in detector technology (CCDs) and software, the first detections of cosmic shear came from four different groups almost simultaneously [17,13,1,9]. This first generation of cosmic shear measurements indicates a low-$\Omega_m$ universe. While this is no suprise, it should be emphasized that this a new method, completely independent of the methods traditionally used to measure $\Omega_m$.

The current generation of cosmic shear surveys will provide more precise constraints, but are all severely limited by telescope time and the small $A\theta^2$ of the existing wide-field telescopes. The Deep Lens Survey (http://dls.bell-labs.com), for example, is planned to take five years to cover five new 4 deg$^2$ fields. Any hope of doing precision cosmology rests on dedicated a high-$A\theta^2$ facility. The design of such a facility is driven by the need to survey large amounts of sky rapidly for control of systematic errors.

Because sources must be resolved for cosmic shear measurements, we must ensure that the size of the $LSST$ PSF is less than the size of galaxies in a broad redshift range. There are three effects to consider: the evolution of galaxy luminosity $L_*(z)$ with redshift, up to redshifts of about 2; cosmological surface brightness dimming proportional to $(1+z)^4$; and the cosmological angular diameter minimum near $z=1$. These three effects produce an observed stalling of the decrease in angular diameter at a given surface brightness as a function of redshift. Thus, galaxies which will be used for $LSST$ weak lens source ellipticity measurement mostly have angular diameters in excess of 0.7 arcsec at the low surface brightnesses typical of the $LSST$ data. The key is to reach faint surface brightness in a time short compared with the timescale of all systematics. The 10-s exposures of $LSST$ will reach sufficient depth to enable control of PSF systematics at unprecedented levels. The co-added images will then go to approximately 30 mag arcsec$^{-2}$ surface brightness, which will yield a high density of resolved source galaxies in the critical $1 < z < 2$ region.

## 4. 3-D MASS TOMOGRAPHY

Cosmic shear is a projected statistic—we can only measure the cumulative effect of all mass at any redshift between source and observer. The development of photometric redshift techniques [3,5] is changing this situation. Now, with images at four or five different wavelengths, source statistical redshifts can be estimated to better than 10% accuracy [18]. The shear shows a strong variation as a function of source redshift, as expected. This variation in turn provides a baryon-unbiased estimate of the lens redshift. The redshifts of mass clusters can be estimated from the shear data alone, without the need for spectroscopy.

3-D mass tomography has now been demonstrated using the combination of weak lensing and photometric redshifts. The $LSST$ will yield three-dimensional mass maps of the universe back to half its current age. More than just maps, these reconstructions will show the evolution of structure in the universe with cosmic time, which will provide a basic check on the foundations of cosmology, and tightly constrain cosmological parameters if our current picture is basically correct. How large a field must be surveyed? Because structure on the scale of $\sim$ 100 Mpc exists, only a survey that samples mass in volumes significantly larger than 100 Mpc on a side will provide a representative measurement of the distribution of mass.

We emphasize that this is a project that cannot be done with existing telescopes. The $LSST$ could reach the required depth of 28–30 mag arcsec$^{-2}$ throughout the wavelength range 0.3-1 micron (needed for color redshift resolution) over a 30,000 deg$^2$ area using five years of dark nights, whereas such a survey on existing 8 m telescopes would take over a century.

Cosmic shear, cluster counts, mass maps, and power spectra are sensitive to different combinations of the cosmological parameters, so we must measure all of these to stringently test cosmological models. Counts of cluster-sized masses, typically $10^{14}$ solar masses and above, are now possible, but the area coverage will be small until $LSST$ comes along. Current mass-selected surveys cover roughly 10 deg$^2$, and when the results of these data are analyzed for mass cluster counts vs redshift in 2006, one might expect to see some constraints in the $\Omega_m$ - $w$ plane at the level of ten percent. However, the wide field of the $LSST$ is critical in getting massive samples of clusters which will provide tight constraints.

Motivated by the discovery that the dark matter in clusters has a non-singular core, there have been recent suggestions that the dark matter may be either decaying or self interacting with hadronic cross sections. To date only two clusters have been studied by this strong lensing inversion. More strong-lensing clusters which form multiple images of a source galaxy must be found. The best way to find these is via a wide-deep lensing survey of the sort $LSST$ will carry out.

## 5. DARK ENERGY

The density and equation of state directly affect the expansion rate and, thereby, the angular diameter distance to cosmic objects. They also affect the evolution of the power spectrum and, thereby, the growth of structure in the universe. For some dark energy candidates, the spatial distribution may be significantly inhomogeneous, which affects the power spectrum and the microwave background anisotropy.

The $LSST$ is a powerful probe of dark energy because it measures a number of properties that depend on different combinations of angular distance and and structure growth. Measurements of the angular distance to nearby clusters constrain the current expansion rate. Combined with measurements of the cosmic microwave background, a measure of the dark energy density can be obtained. The number density of clusters (mass selected via weak-lensing) as a function of redshift is dependent on both the volume-redshift relation and the growth function. These in turn are sensitive to the amount of dark matter $\Omega_m$, the amount of dark energy $\Omega_E$, and the dark energy equation of state $w$.

To solve the general problem of determining the density and equation of state of dark energy without prior assumptions is challenging and requires a suite of independent measures. This points to the advantage of $LSST$: by measuring comic shear as a function of redshift, the degeneracy between the normalization of the mass powerspectrum $\sigma_8$ and $\Omega_m$ seen at low redshift is broken. By surveying the volume number density of massive clusters vs redshift the comoving volume can be measured over a range of redshifts where the effects of dark energy are maximal. $LSST$ simultaneously performs a number of independent tests. How these tests combine among themselves and with other cosmological tests to provide constraints on dark energy without prior assumption is an important subject for simulation.

### 5.1. Measuring the dark energy density

The first major contribution of $LSST$ will be the independent measurement of $\Omega_E$ through cosmic shear and projected cluster counts as described above. Is the result from SNe of non-zero $\Lambda$ correct? Current deep pencil beam lensing surveys covering 30-100 square degrees may answer this question to 10% precision, and if we assume a prior for several cosmological parameters, even better precision. But the right way is not to assume we have the right theory, but to make sufficient weak lens observations covering a large volume such that degeneracies between parameters are broken and we arrive at unique solutions. Such a result will come out of the $LSST$ weak lens program. Figure 2 shows were we can be after 5 years of $LSST$ for the determination of the density of dark energy. Up to a billion source galaxies will be used, degeneracies between parameters will be broken, and the data analysis will marginalize over all cosmological parameters.

Of even greater interest, assuming that the dark energy density is found to be non-zero, is the physical nature of the dark energy. A first step in probing its nature is the measurement of the dark energy equation of state $w$.

### 5.2. LSST dark energy cluster lens survey

The mass cluster counts are sensitive to dark energy through its effects on the growth of density perturbations and the comoving volume element as a function of redshift. This has been noted previously and has prompted several groups to plan large surveys using the Sunyaev-Zel'dovich effect, X-ray emission from clusters, and OII emission from galaxies (DEEP). Compared to these methods, weak lensing has the advantage that it probes the mass directly, independent of assumptions about gas dynamics. It is probable that 20% of clusters detected via weak lens shear are unvirialized [15], and it is important when comparing with N-body predictions to do a mass survey rather than a survey biased to radiation. Compared with cosmic shear, cluster weak lensing has the advantage that it is in the relatively high S/N regime, i.e., it does not depend critically on being able to control shear systematics at the 0.001 level. This mass cluster counting is a key project in the $LSST$ pilot survey called the Deep Lens Survey which will survey 28 square degrees to 26 magnitude in four bands. The results of this pilot survey now under way, will however constrain $w$ to about 10%.

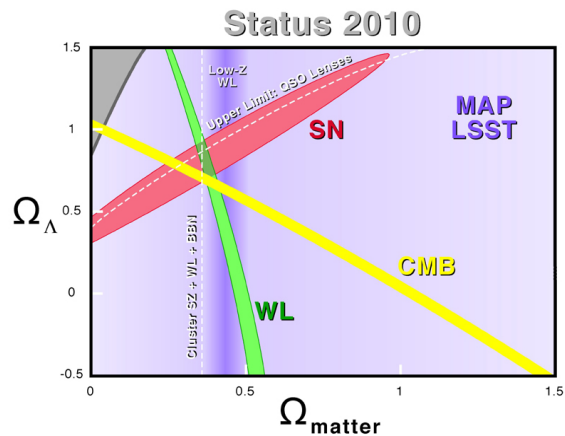

Figure 2. An example of the possible status for the determination of the dark energy density in the decade starting in 2010. Weak lens $LSST$ observations of source galaxies at high redshift will define the thin WL confidence band. $LSST$ weak lens observations of low-z source galaxies will define the vertical band. Will the SN and WL solutions cross the CMB curve at the same point? This is one example of a test of the underlying theory.

The figure of merit for determining cosmological parameters via mass tomography is a combination of depth and area coverage, with a premium on the latter. $LSST$ will resolve a higher density of background source galaxies and will cover a far larger area of 30,000 square degrees. Combined with the corresponding larger redshift range for the 3-D tomographic mass study, $LSST$ will cover an un-precedented volume: over 100,000 mass clusters, distributed over the critical redshift range 0.2 - 1 for maximum sensitivity of the effects of dark energy. This method provides a dark energy probe that is complementary to SNe Ia: the cosmological parameter degeneracies are nearly orthogonal, and the systematic errors are completely different.

What precision do we expect today on $w$ from CMB, SN, and galaxy power spectrum, in the absence of lensing data? Combining current data for CMB, SN, and $P_{\rm galaxy}(k)$ data, a $3\sigma$ accuracy of 30% is in principle obtained [2]. A mass prior or including mass tomographic data would increase this accuracy significantly.

### 5.3. Strong constraints on w from 3-D tomography

Cosmological parameter estimation from shear power spectra benefits from 3-D mass tomographic reconstruction of the mass power spectrum vs cosmic time. If we increase the number of redshift bins for the source galaxies in an $LSST$ weak lens survey to several bins covering the z= 0.2 - 1.5 range, the constraints on all cosmological parameters are improved. For example, consider the predicted error contours in the $w$-$\Omega_E$ plane from NASA's CMB $MAP$ satellite, and the $LSST$ weak lensing survey – with and without source photometric redshift information. This $LSST$ 3-D weak lens tomographic survey will produce the enhanced constraint on $w$ shown in Figure 3.

### 5.4. Constraints on $w$ from CMB and counts of mass clusters out to $z \sim 1$

What can we learn from wide-deep weak lens surveys of mass clusters, when combined with data from the cosmic microwave background (CMB)? The CMB data yield a precise value for $\Omega_m + \Omega_E$. Combined with the cosmic shear data and/or the mass cluster data from 3-D tomography, CMB data removes the degeneracy in $\Omega_m$-$w$ space. The largest contributor to the mass cluster number-redshift relation in the redshift range where dark energy effects dominate ($z < 1$) is the comoving volume-redshift relation. At higher redshift structure development begins to dominate. Generally, cluster redshift distributions are far more sensitive to cosmological parameters than are luminosity distances. Clearly we must compare the observed mass cluster counts to full N-body simula-

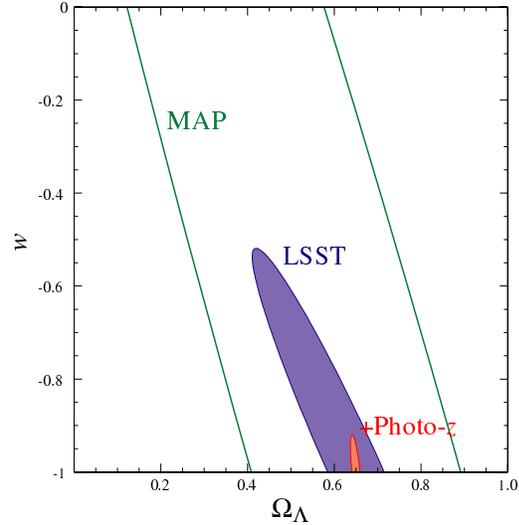

Figure 3. Error ellipses for the $\Omega_E$-$w$ plane (dark energy density vs. equation of state) for a 1000 square degree survey with $LSST$, only 3% of the full survey. Photometric redshift information with $LSST$ ("Photo-z") leads to a substantial gain in the precision due to tomographic breaking of parameter degeneracies. By surveying 30,000 square degrees, $LSST$ will test for cosmic variance. (Hu 2001)

tions for various candidate XCDM models.

The most reliable approach is to generate mock data: simulate the growth of stuctures within the context of some XCDM model, shine light through it, and analyze the reulting shear maps of background galaxies. We have undertaken such a study for the twin purposes of comparing with the results from the DLS and predicting the accuracy of the $LSST$ dark energy key project. Figure 4 shows a mass map for a LCDM simulated DLS 1-degree (diagonal) field. We have repeated this multiple times and for the vastly improved noise, resolution, and coverage of the $LSST$. Since we are comparing the number of detected clusters in the observations directly to numerical simulations, projection effects and chance alignments are explicitly included. This circumvents many of the biases associated with comparing the number of clusters to analytical predictions [16]. An $LSST$ weak lens mass cluster survey should see all clusters which have a critical kappa value and with sources at $z = \sim 2$.

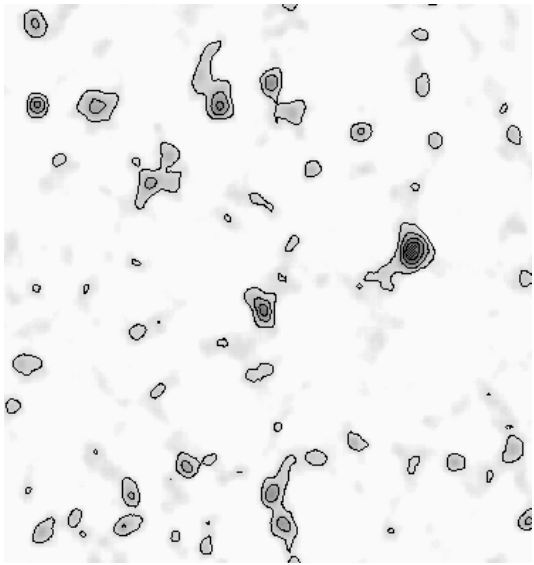

Figure 4. An LCDM N-body simulation for a 2 × 2 degree DLS field. Galaxies to z=1.5 and to 25 mag were used as sources for the shear measurement. The resulting projected mass map is shown. The statistics of these N-body mock data are very similar to those of the current Deep Lens Survey. $LSST$ will go much fainter and will cover 1000 times the area.

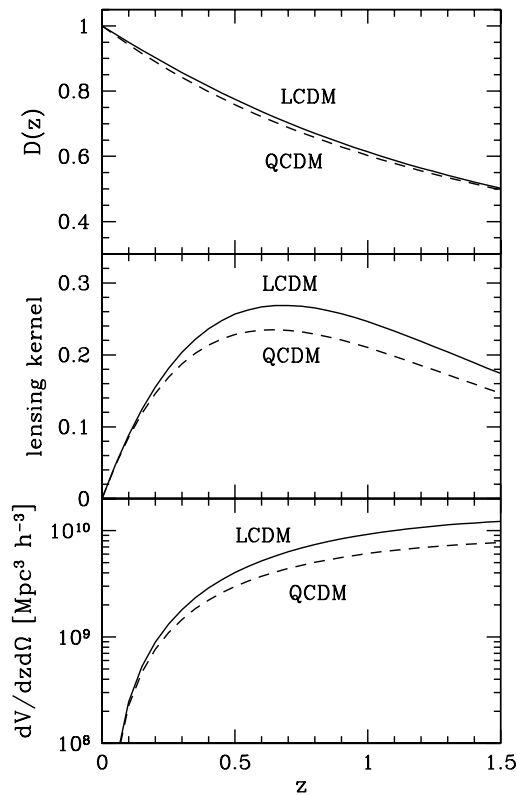

Figure 5. Using a flat geometry (from CMB) two cosmologies with different $w$ are shown here. The number of mass clusters depends on the growth of structure (top), the lensing kernel (middle), and comoving volume (bottom).

To test the sensitivity of $LSST$ to dark energy equation of state, it is useful to consider two cosmological models (with different w) that have the *same* CMB spectrum. The two models are: LCDM with $\Omega_m = .28$, $\Omega_\Lambda = .71$, $h = .68$, $\sigma_8 = .84$, and $w = -1$; and QCDM with $\Omega_m = .40$, $\Omega_Q = .6$, $h = .56$, $\sigma_8 = .73$, and $w = -2/3$. Although these two models are indistinguishable even with the MAP CMB data [8], they are clearly distinguishable with weak lens mass tomography. The observed number of clusters depends on the lensing kernel and volume surveyed. Figure 5 shows how, for two models with identical CMB anisotropy and nearly identical growth of structure, two cosmologies with different $w$ may be distinguished based on mass cluster counts as a function of redshift [4]. Weak lensing is free of the usual standard candle or standard meter stick systematics, since only angles and redshifts are measured.

Figure 6 shows the normalized redshift distributions for clusters detected in an LSST lensing survey for the two models (LCDM & QCDM) with degenerate CMB anistropy spectra, but different values of $w$. These curves assume that the completeness of the survey for clusters with $M > 2 \times 10^{14} M_\odot$ is given by the lensing kernel in Figure 5 normalized to unity, and were obtained by taking the product of this normalized kernel and the cumulative mass function of dark halos. For a more detailed analysis using N-body simulations see [4]. As an indication of what will be found in these N-body simulations, we plot the modified Press-Schechter mass function lens kernel product as a normalized probability vs redshift. It is clear from Figure 6 that the redshift distribution of clusters detected by LSST can be used to distinguish the two models. These theoretical models differ sufficiently that a sample of 10,000 clusters could give a statistical precision on $w$ of about one percent. The dominant error will likely be the error in determination of the mass-redshift selection function, although comparing plots of normalized probability vs redshift minimizes that systematic error. Any such systematic will likely be a function of cluster mass, and could be investigated by comparing the mass function at fixed redshift with N-body results. Cosmic variance should also be consistent with the best fit theoretical N-body simulation. Finally, sample variance will enter at the one percent level as well. All such effects will vary across independent volumes and different redshifts. Thus, systematics could be tested if one cuts ten such samples in various ways. This total sample of 100,000 clusters above $2 \times 10^{14} M_\odot$ will be obtained with $LSST$. By measuring mass directly in a variety of samples we expect to control mass systematics to support a percent level determination of $w$.

Counting only clusters above of $2 \times 10^{14} M_\odot$, $LSST$ will produce a sample of over 100,000 mass clusters with accurate redshifts and calibrated masses, which will determine the cluster number count distribution with redshift to high precision. Comparison of the shape of the observed distribution function to N-body simulations will constrain $w$ to about one percent with control of systematics at that level.

### 5.5. Prerequisites and outlook

Mass finder algorithms should be tuned prior to $LSST$, via the current lens surveys like the Deep Lens Survey, so that the lensing pipeline requirements are known before $LSST$ begins operation. Similarly, lensing and photometric pipeline software work should begin soon, well in advance of operations. For weak-lens tests of dark energy/matter which rely on the evolution of the non-linear part of the mass power spectrum, more complete and larger N-body simulations for a wide range of $\Omega_E$ and $w(z)$ must be carried out. This will happen before $LSST$ shear data come in, but it is important to do this work now for key program planning. For the mass cluster counting constraint on $w$ we rely on larger shear and lower redshift, so uncertainties in the power spectrum and growth of structure at high z are less important.

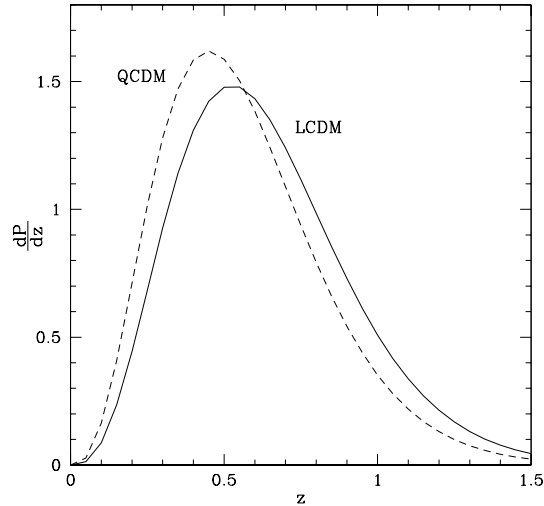

Figure 6. The power of mass cluster counts to discriminate different $w$. The normalized probability of cluster shear detection above a mass threshold is plotted vs redshift. Only the flat geometry (from CMB) was assumed. Two cosmologies are shown: LCDM ($w = -1$) and QCDM ($w = -2/3$). Precision constraints on $w$ will result from the statistics of 100,000 mass clusters measured with $LSST$ and with the unprecedented control of systematics.

For the $LSST$ SN program, a better understanding of the reliability of SNe as standard candles will be necessary, particularly with respect to possible evolution effects. Considerable progress on this will have been made by the time $LSST$ begins operations. Ultimately, an orbiting IR imaging wide-field imaging facility would be helpful in constraining these SN systematics.

There is also some hope that 3-D tomography will constrain the rate of change of $w$. With precision results for $\Omega_m + \Omega_E$ from MAP, and an $\Omega_m$ prior from an array of observations, mass tomography over 30,000 square degrees could measure $w'$ to a few percent[7]. Within a decade there will be complementary observations of the geometry of the universe and the effects of dark matter and dark energy. Will any of the current models survive? These observations range from Sunyaev-Zel'dovich effect to mass tomography to SN tests to CMB precision asisotropy. They probe the volume-redshift relation, the mass power spectrum, and galaxy cluster abundance in different ways. They will jointly lead to a precision constraint on the expansion history of the universe bracketing the apparent mass dominated to dark energy dominated transition.